# DNA based Network Model and Blockchain

A.M. El-Edkawy, M.A. El-Dosuky and Taher Hamza

Computer Science Department, Faculty of Computers and Information, Mansoura University, Egypt

amr.eledkawy@mans.edu.eg, mouh_sal_010@mans.edu.eg, taher_hamza@yahoo.com

**Abstract:** *Biological cells can transmit, process and receive chemically encoded data in the same way as network devices transmit, process, and receive digitally encoded data. Communication protocols have led to the rapid development of computer networks. Therefore, we need to develop communication protocols for biological cell networks, which will lead to significant development, especially in medical applications where surgery or delivery of drugs can be performed using nanoscale devices. Blockchain is a peer-to-peer network that contains a series of clusters to make a valid and secure transaction. Blockhain technology is used in many areas such as e-commerce, public services, security, finance, Internet stuff, etc. Although blockchain has a major impact on Internet technology, it suffers from time problems and scalability. DNA computing is the execution of computations using natural molecules, especially DNA. DNA gaps above silicon because of massive parallelism, size and storage density. In this paper, biological cells and DNA are used to create the necessary protocols for the networks to be used in the performance of the cell-based communication system. The proposed hybrid solution involves DNA as well as calculated on an enzymatic basis, where each contributes to the function of a given protocol. Also a correspondence between blockchain and DNA is proposed that can be utilized to create DNA based blockchain.*

## 1. Introduction

Biological cells can transmit, process and receive chemically encoded data in the same way as network devices transmit, process, and receive digitally encoded data [1]. Biological cells resemble biological devices that have biological interfaces (gap junctions, receptors), biological memory stores (nucleic acids), computing processes (enzyme pathways, and regulatory networks). The data are chemically encoded from a series of bio-chemical DNA bases. The modern technologies available to install custom DNA threads and custom software tools available today facilitated the development of DNA-based solutions.

Communication protocols have led to the rapid development of computer networks. Therefore, we need to develop communication protocols for biological cell networks, which will lead to significant development, especially in medical applications where surgery or delivery of drugs can be performed using nanoscale devices.

At the end of 2008, the Bitcoin [2] was developed and first developed blockchain was used. At present, many organizations are considering using blockchain for use cases other than encryption currency. The absence of the central administrator in the blockchain has many advantages to the companies because it reduces costs and makes the procedure simple and fast. The blockchain security is very strong as only the connected hosts can access the contract data.The Bitcoin online trading system provides portfolio management services where the wallet stores a pair of private and public keys as well as storing transactions in the system. The wallet also stores encrypted user preferences. The key pair is used to send received Bitcoin where public keys are given to senders to identify recipients while special keys are used to sign the transaction and to ensure exchange of encrypted currency [3]. Blockchains are a kind of distributed ledger techniques. Blockchain is a peer-to-peer network that contains a series of clusters to make a valid and secure transaction. Because the ban chains are peer-to-peer networks, they do not have any centralized control, and data quality is guaranteed by replication methods and computational trust. Blockchain data is linked to linked and locked blocks using decryption and encryption techniques [3]. This string is not changeable. Blocks have the following characteristics: Decentralization, Auditability, Stability and anonymity. Blockchain has been used in many leading companies such as IBM and Microsoft. Blockhain software is used in many areas such as e-commerce, public services, security, finance, Internet stuff, etc.

Although blockchain has a major impact on Internet technology, it faces a number of challenges. First, the problem of scalability, where the volume of the mass of the maximum volume of the volume of 1 megabyte, and take the extraction of one block about 10 minutes, although the size of the block is small, many transactions may be delayed because the miners responsible for the creation of the following block block transactions It has a higher fee. Thus, Bitcoin blockchain is limited to 7 transactions per second rete, resulting in a lack of real-time processing of millions of transactions. The larger the size of the pieces, the slower the spread in the blockchain network and the creation of branches. One of the basic concepts in blockchain is the consensus concept which means that all transactions are replicated in each peer in the grid, thus checking the block requires a complete blockchain that needs higher volumes. Also, it may take less than one hour for validation to be required, for example, in slavery services, especially if it is used offline in non-Internet stores. So the blockchain suffers from time problems and scalability. The transaction increases every day and shows that the blockchain is heavy. The storage volume in the Bitcoin block now exceeds 100 GB [3].

In bitcoin blockchain, the minor adds new bitcoin currencies using a special program to explore the bitcoin coefficients that must be verified [2]. Verification requires solving complex math problems that require higher processing power [3]. Also, current compatibility algorithms such as proof work validation require higher processing capacity. So the power is a difficult problem in the blockchain. Many special computer components that are mainly provided by China have an important role in current blockchain technology. Silicon chips represent the core of computers for more than 50 years. According to Moore's law, the amount of electronic gadgets on the chip multiplied every 18 months. Many predicted that Moore's law would soon come to an end, in the light of physical speed and curtailing the restrictions on silicon chips so that there is no Moore's Law [4].

DNA computing is the execution of computations using natural molecules, especially DNA, rather than silicon. DNA computing is a new and attractive development as it is multidisciplinary between molecular biology and computer science. It has recently evolved not only as an attractive information processing innovation, but also as a catalyst for the exchange of learning between biology, nanotechnology and information processing. This area of research can change our understanding of the hypothesis and developments in computing.

DNA gaps above silicon consist of the following reasons [5]:
- Massive parallelism: calculations can be performed simultaneously, rather than sequencing in silicon.
- Size: DNA computers are much smaller than silicon devices.
- Storage density: More data can be placed in a similar amount of space.

. In this paper, biological cells and DNA are used to create the necessary protocols for the networks to be used in the performance of the cell-based communication system. The proposed hybrid solution involves DNA as well as calculated on an enzymatic basis, where each contributes to the function of a given protocol. Also a correspondence between blockchain and DNA is proposed that can be utilized to create DNA based blockchain.

In the study, we also explained how blockchain was applied using DNA techniques that would solve the scalability problem of existing block chains. The paper is designed as follows: Section 2 reviews related work. Section 3 introduces the proposed cell-based communication platform. Section 4 shows how to implement a blockchain using DNA computing techniques. Finally, section 5 presents conclusions and future actions.

## 2. Related Work

Molecular communication uses encrypted molecules as information carriers for the engineering of biochemical-based communication systems. In [6], Moritani et al. Define the molecular communication interface that uses inline vesicles with articular junction proteins to transfer the coding molecules of messages. Gap junctions form channels of communication between cells and vesicles that allow small molecules such as ions, metabolites and small nucleotides to diffuse from the cytoplasm to the vesicle and vice versa. Vesicles act as signal carriers, and signal molecules are spread between transmitter and receiver transmitters. The selectivity and permeability of the gap-crossing channel is influenced by various factors such as the connexin phosphorylation [4] and the environmental pH and temperature. We will later describe how external control is used in selectivity and permeability characteristics to perform functions such as link switching, a form of guidance found in traditional data network devices [4].

In [7], a self-programmable, self-contained mechanism is made entirely of biomolecules. The design consists of a long DNA input molecule that is frequently treated by a restriction enzyme. "Short" DNA molecules control the action of the restriction enzyme precisely. This concept forms the basis of a nanoscale arithmetic machine that diagnoses the disease and disseminates treatment molecules based on several inputs that refer to diseases [8]. It is expected that more sophisticated machines such as stack automation and applications will be developed to encode programmable DNA molecules using similar techniques. In [9], Liu et al expanded the molecular mechanism presented in [8] to the DNA-based automatic killer design, which can release cytotoxic molecules that spread to neighboring cells via the interstitial channels.

However, there are many challenges in creating complex circuits within the body such as chemical heterogeneity, uniformity and predictability [11]. In [12] a series of interrelated logical gates based on similar enzyme interactions that are successfully regulated. To achieve unity, each subunit of logic must use splitting mechanisms, for example distinct chemical types that have been designated to prevent intrinsic chemical interference between the gates or the specificity of scaffold particles [13]. These mechanisms can provide arithmetic functions to support the nanoscale account of networked devices.

## 3. Proposed DNA based Network Model

The following figure represents the Data Flow in the OSI Model [14].

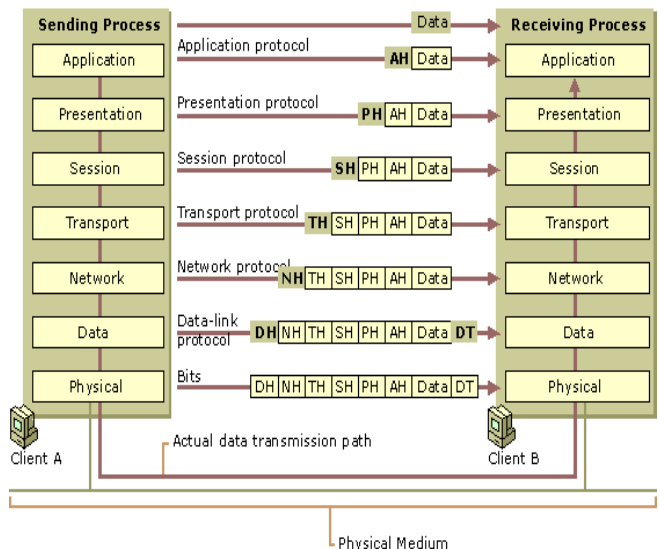

Fig. 1. Data Flow in the OSI Model [15]

In the following paragraphs represents how to represent that data flow model using the proposed cell-based communication protocol.

The data in the system would be represented as sequence of DNA nucleotides ACGT.

- Application layer

The application layer acts as a window for users and application processes to access network services. This layer contains a variety of commonly needed functions: resource sharing and device redirection, remote file access, remote printer access, process communication, network management, directory services, e-mail (eg mail), virtual network terminals [14].

The application layer attaches an application header and then passes the frame to the presentation layer. The application header would be specific for each application and then would be added to the data by using the ligation process to attach the data to the application header.

- Presentation layer

The presentation layer is used for transform data in various ways, if necessary, such as by translating it and adding a header. It gives the result to the session layer.

But this responsibility of the presentation layer is not required as all data would be from raw in the DNA form. But if the data are not represented in the DNA form. It can be transformed using the codons table where every codon can be viewed as correspondence from another translating table from different kinds of data in order to translate it to the codons form.

Table 1. Amino acids and codons

| Genetic Code: Codon Translations for Amino Acid Residues ||||||||
|---|---|---|---|---|---|---|---|
| UUU | F | UCU | S | UAU | Y | UGU | C |
| UUC | F | UCC | S | UAC | Y | UGC | C |
| UUA | L | UCA | S | UAA | stop | UGA | stop |
| UUG | L | UCG | S | UAG | stop | UGG | W |
| CUU | L | CCU | P | CAU | H | CGU | R |
| CUC | L | CCC | P | CAC | H | CGC | R |
| CUA | L | CCA | P | CAA | Q | CGA | R |
| CUG | L | CCG | P | CAG | Q | CGG | R |
| AUU | I | ACU | T | AAU | N | AGU | S |
| AUC | I | ACC | T | AAC | N | AGC | S |
| AUA | I | ACA | T | AAA | K | AGA | R |
| AUG | M | ACG | T | AAG | K | AGG | R |
| GUU | V | GCU | A | GAU | D | GGU | G |
| GUC | V | GCC | A | GAC | D | GGC | G |
| GUA | V | GCA | A | GAA | E | GGA | G |
| GUG | V | GCG | A | GAG | E | GGG | G |

The presentation header then can be added using the ligation process

- Session layer

The session layer allows session establishment between processes running on different stations. The session layer header can be added using the ligation process.

- Transport layer

The transport layer ensures that messages are delivered error-free, sequentially, without any loss or duplication. It relieves protocols of the upper class of any concern about the transfer of data between them and their peers [14].

The message segmentation in the transport layer can be done in the message by using of cutting enzymes. We can make the encoding and decoding way based on the method in [6] that uses "protector strands" to control the operation of an enzyme based state machine by separating the constituent DNA strands of message molecules.

Also messages require error detection and correction. Invariably, errors will occur in the encoding and transmission process of DNA molecules due to the imprecise nature of the associated complex biochemical reactions [1]. By including redundancy in the encoding process, error correction mechanisms can be incorporated into the decoding process.

The network header then can be added using the ligation process.

- Network layer

The network layer controls the operation of the subnet, and specifies the actual path that data should follow based on network conditions, service priority, and other factors [14].

Title coding can be performed similarly to the model proposed in [9] and [16] to create a DNA-dependent mechanism that produces contact molecules. Each message is encrypted as a unique string of nucleotide bases as described in [7]. Similar to the techniques used in [9] and [8], control of the calculation is performed by releasing molecules (such as mRNA) that selectively activate the "base" molecules in the DNA. It is suggested that cleaved molecules during calculation can provide input to other parallel mathematical functions. The cleaved ssDNA message particles are released in the cytosol and provide input to the control function in the molecular interface of the grid layer [16]. In theory, this mechanism can be extended to encrypt many unique address locations and any number of messages during the calculation.

• Data Link layer

The data link layer provides error-free transmission of data frames from node to node across the physical layer, allowing the layers on them to actually assume error-free transmission over the link [14].

Methods used above for error detection and correction can be used here again and headers also can be added using the ligation process.

• Physical Layer

The physical layer, the lowest layer in the OSI model, is concerned with the transmission and reception of the raw raw bit stream through a physical medium. Describes electrical / optical, mechanical and functional interfaces to the physical medium, and carries signals to all upper layers [14].

The transition of the physical layer depends on solutions by [1, 17] for molecular communication. The interface selection can be achieved using the "real world" application for logical repeating architecture as described in [17]. The address encoding molecule is entered into the switching circuit, which then fires / changes the chemical signal corresponding to "turning" the message to the correct interface. In the case of the intersection of the gap, the output of the enzyme-based circuit will control the permeability of the gap-crossing channels. The permeability of the gap junction with connexin phosphorylation is affected by specific phosphorescence reagents, whose concentration is controlled by the switching circuit. Thus, the circuit can be used to turn on and off each molecular interface effectively by controlling the phosphorylation in the connexins of the intersection of the gaps. This in turn will allow the encrypted message to be transmitted via only one link (or multiple links if using multicast). Using this technique, several connections can be controlled simultaneously at the same time through fragmented enzymatic functions.

## 4. DNA based Blockchain

There are 11 global DNA collection centers have been opened around the world. We have understood our findings on how transitions in our ancestors have shaped our DNA over the ages, making it possible for today's scientists to follow our origins 200,000 years ago. The fact that some animals have access to a genetically transmitted collective memory tells them, for example, that some species are natural enemies and that they are very dangerous to prove that there is a data warehouse and that it holds some memories [18].

Mutation information is similar to blockchain technology. Mutation information is distributed and stored to all deceased subjects, possibly millions of people on-line, making them as safe as blockchain information, and now know how to access them.

The similarity between Blockchain and DNA is summarized in the following table:

Table 2: Similarities between Blockchain and DNA

| Blockchain | DNA |
|---|---|
| wallet of Bitcoin (The wallet of Bitcoin has a single Blockchain with the information) | Cells of our body (Cells has a single DNA with information) |
| Blockchain (Bitcoin has a lots of wallet about s and the wallets have the same Blockchain.) | DNA (The human body contains 15 trillion cells and the cells of body have the same DNA) |
| addresses in the coin wallet. (Each wallets of bitcoin has his own address. And they can express the balance of each wallets.) | Expressed proteins Cells in a particular body, for example the heart or the skin, will express the specific proteins via the mRNA by recognizing the respective positions of body. That is, the cells are expressed according to its location in the body.) |
| For new wallet to create, it must first replicate its Blockchain | For the cells to divide, it must first replicate its DNA |
| The content of DNA is not be erased. | The content of Blockchain is not erased after six conformation in bitcoin. |
| The DNA has sometimes a mutation with low probability during replication. That is, the DNA of all the body is not the same DNA. | The Blockchain has sometimes multiple chains during forking. |

The difference between the two, the blockchain record is increased when the transport. But the DNA of cells does not change.

Thus, DNA resembles a bitcoin block. They both store duplicate data, and DNA in life forms are spread all over the planet, between chains in computers that occupy Pitcuin throughout the Internet. "Data" stored is the accumulation of transactions in the real world (financial transactions versus genetic transactions, or mutation).

Biological evolution is similar to the buildup of blocks and the processes of bitquin. The data is slowly changed over time: the evolution of the gene changes and the additional parameters and blocks change the blockchain. Both blockchain (always) and DNA (sometimes) become more complex over time.

The working guide is survival of the fittest. Any single replica can be modified, altered, or artificially altered - for Bitcoin by manually editing the local blockchain, DNA through genetic engineering, breeding or mutation. But in both cases, for local change to become global, changes to the data store must undergo a very difficult test. In the case of DNA, DNA must produce a viable living organism and multiply in the real world. In the case of Bitcoin, the rest of the network must accept the cluster as legitimate by confirming it to Proof of Work. If local changes are easily replicated by the world population without such a test, there will be chaos. In life, survival is really proof of action.

Blockchain fork are closer to the speciation. Altcoins are more like biological orders.

The rogue nodes in the Betquin like failed DNA mutations. The method of system design means that they do not pass the test, so they do not spread and do not last in general.

Bitcoin asserts that it resembles "good" DNA mutations. These mutations pass the test, spread across the network (or through the DNA of the population).

Both depend on global infrastructure. Bitcoin requires a Betquin network and the Internet. DNA requires a network of life on earth to provide everything it needs.

It is similar to predation among species that take Altcoins market share from each other. More market share means having a more active bloc / more breeding and chances of a boom.

Finally, both DNA and Blockchain are in essence the epitome of truly epic tales. In the case of DNA, it is the story of life itself, from the first beginnings, to every layer of evolution and origin. In the blockchain, every transaction in the history of Bitcoin has been immortal and wonderful.

## 5. Conclusion

In this paper, biological cells and DNA are used to create the necessary protocols for the networks to be used in the performance of the cell-based communication system. The proposed hybrid solution involves DNA as well as calculated on an enzymatic basis, where each contributes to the function of a given protocol. Also a correspondence between blockchain and DNA is proposed that can be utilized to create DNA based blockchain.